\documentclass[sigplan,screen,10pt]{acmart}
\renewcommand\footnotetextcopyrightpermission[1]{}

\usepackage{subfigure}
\usepackage{xurl}

\begin{document}
\author{Ragini Gupta}
\email{raginig2@illinois.edu}

\affiliation{%
\institution{University of Illinois, Urbana-Champaign}
}

\author{Akash Mittal}
\email{akashm3@illinois.edu}
\affiliation{%
\institution{University of Illinois, Urbana-Champaign}
}

\title{MuTable (Music Table): Turn any surface into musical instrument}

% \author{Haoran Qiu, Beitong Tian, Ragini Gupta}
% \affiliation{
  % \institution{University of Illinois at Urbana-Champaign}
%}
% \email{{, , raginig2}@illinois.edu}

\maketitle

\section{Abstract}
With the rise in pervasive computing solutions, interactive surfaces have gained a large popularity across multi-application domains including smart boards for education, touch-enabled kiosks for smart retail and smart mirrors for smart homes. Despite the increased popularity of such interactive surfaces, existing platforms are mostly limited to custom built surfaces with attached sensors and hardware, that are expensive and require  complicated design considerations. To address this, we design a low-cost, intuitive system called \textit{MuTable} that repurposes any flat surface (such as table tops) into a live musical instrument. This provides a unique, close to real-time instrument playing experience to the user to play any type of musical instrument. This is achieved by projecting the instrument's shape on any tangible surface, sensor calibration, user taps detection, tap position identification, and associated sound generation. We demonstrate the performance of our working system by reporting an accuracy of 83\% for detecting softer taps, 100\% accuracy for detecting the regular taps, and a precision of 95.7\% for estimating hand location.     
\section{Introduction}
\label{sec:introduction}
With the recent advancement in new display technologies (such as video projectors, plasma screens, LED matrices, multi-monitor video walls, electronic links), interactive surfaces have become a pervasive solution for various application domains. Often these display surfaces are passive output devices and despite being fully interactive via a touch screen, they lack input capabilities apart from an external keyboard, mouse or a remote control. Moreover, these interactive surfaces are  limited to only small sized display screens or large projection walls. Several researches have used computer vision techniques with large projection walls in areas of Augmented Reality \cite{surface,AR1}  and interactive dance \cite{interactivedance,InteractiveDance2} to monitor user activity in an extended room, however, they do not explore quantifying precise human gestures and movements at the surface of the screen.  One such application that can leverage the use of interactive surfaces is, musical instruments. 

Learning any musical instrument requires regular practice, time and effort in order to be adequately skilled. However, musical instruments such as piano, conga drums, bass etc. are expensive, heavy and bulky in nature that limits their availability and ease of accessibility across a wider population, thus limiting their instrument learning process. Therefore, as an alternative solution video tutorials, smartphone music apps or music games gain a large popularity among the users.   
Additionally, the process of playing these instruments is a tedious, repetitive process that requires precise eye-hand coordination. This makes it a challenging task for individuals with motor disabilities particularly those with Parkinson's disease to enjoy a seamless instrument playing experience.

Using techniques such as computer vision, hand gesture and movement tracking against the surface projection, one can design a tangible, low-cost, easily accessible interactive surface for playing musical instrument. It is worth mentioning that we limit the scope of our proposed system to flat horizontal surfaces to provide an evocative instrument playing experience to the users.   

\begin{comment}{ Most of these big musical instruments also require large space and regular maintenance which is a major bottleneck for buying and managing these instruments.  }
\end{comment}
Previous technology innovations have tried to address some of these limitations by designing musical instruments using augmented and virtual reality \cite{r1,chromacord}, musical instrument apps \cite{musicaa, app}, and 3D printed instruments, but the experience is far from reality. Existing computer assisted instrument augmentation have also tried to improve one or more features of the instruments such as automated tuning \cite{property1}, sensors to improve sound quality \cite{hand} , tracking human motion while playing instrument \cite{pianist1}, etc. However, these technologies are only suited for improving the hardware design, sound, aesthetics and usability of existing physical instruments. 

Therefore, in this work, we aim to create an environment where one can have a virtual instrument playing experience by transforming any flat surface into an intuitive, interactive, tangible instrument interface. 
This prototype can empower the users to practise playing a musical instrument seamlessly without the need for any instrument or a pre-designed surface. The virtual musical instrument system will consist of a basic setup with a
camera, a pair of wearable band and a projector. The system provides a low-cost, portable solution for a unique virtual instrument playing experience similar to playing a real instrument without any exposure to digital screens. Since there is limited focus on virtual musical instruments for drums in the existing literature, therefore, in this paper we emphasize on designing, implementing and testing a new virtual interactive surface for playing a specific type of musical drums, i.e. the conga drums.

A proposed system overview is illustrated in Fig. 1. The architecture consists of an IMU sensor equipped wrist band on the user's hand to detect the tap intensity and to track the hand movement of the user.  A projector will be installed at an elevated platform to project the drum's area on a given passive surface, such as table top. A camera will be hosted along with the projector to perform fast gesture recognition coupled with hand localization on the projected surface in real-time. Combined processing using IMU sensor data and images will be used for spatial sound synthesis which will be played aloud via a connected speaker.

\begin{figure}
\centering
\includegraphics[width=\linewidth] {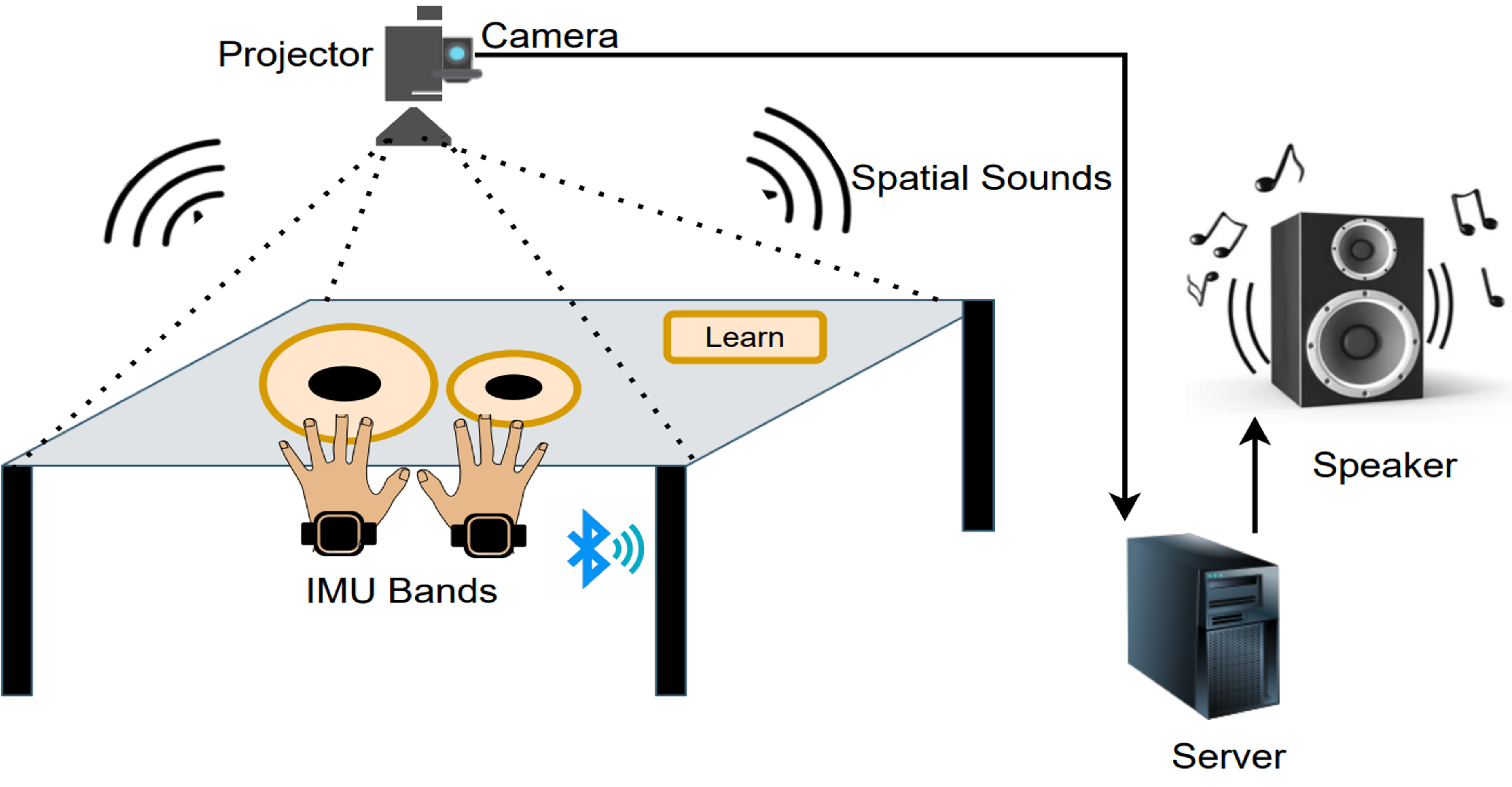}
\caption{System Overview
}
\label{fig:Architecture}
\end{figure}

Given the multi-modal input data into the MuTable system (such as the IMU data, depth, images), one of the major challenge is the intelligent fusion of different modalities in real-time such that realistic sounds can be generated depending on the user's tap position, tap intensity, and instrument shape. 

Other significant challenges of the proposed system include first, low sound delay from the time a tap is detected using the wearable band till the sound is produced and second, high accuracy for hand localization and tap detection on the projected surface. In order to address these challenges, we will incorporate different approaches such as adaptive depth camera calibration, self-calibration of IMU sensors on wearable bands, optimized data computations from live camera streams, discretization and sensor thresholding techniques. Each of these methods will be discussed in detail in the later sections of the paper. 

We evaluate the performance of MuTable system for end-to-end latency, tap detection accuracy and hand localization accuracy. Our experimental results show that we achieve an end-to-end latency of 100.3 ms. Additionally, the system achieves the tap detection accuracy of 83\% for softer taps and 100\% accuracy for regular taps. Precision for the estimated tap position is reported to be 95.7\%.

% talk about the trends of moving to serverless computing paradigm or "FaaS"
%\subsection{Background}

% Motivation+Concrete Problem Statement (need not be formal, only concrete).
% talk about the motivation for this project and for the following problem statement
%\subsection{Motivation}

% talk bout the problem statement
%\subsection{Problem Statement}
\section{Related Work}
\label{sec:related_work}

\subsection{Interactive Surfaces}
The design of interactive surfaces in mobile computing has been a research focus in the recent years. Subsequently, the researchers have developed technologies  such as ScratchInput \cite{scratchInput}, Skinput \cite{skinput}, LumiWatch \cite{lumiwatch} to project small devices' screens and displays on passive surfaces such as desks, walls and even human body. These projected surfaces can then be used as a kind of input surfaces for mobile hand-held computing devices that have tiny displays, buttons, and keyboards. Projection of such interactive surfaces are only limited to computational device displays and I/O peripherals. Applicability of such interactive surfaces for musical instruments is very limited. One such work is called BendAble \cite{bendable}, which is a multi-sensory interactive interface for autistic children to practise coordination by producing sounds. The rich multi-sensory stimuli can improve the overall learning experience for the children as they continue to obtain haptic sensations from the surface to self-regulate their movement strength and timing. However, it does not leverage the use of any physical surface and is only limited to using a pre-designed interactive canvas-like surface attached with multiple sensors. 
\subsection{AR/VR platforms for instrument learning}
Augmentation of musical instruments, particularly piano, has been done in the past to emulate musical instruments to provide a new learning and playing experience. For example, authors in \cite{m1} survey the existing augment reality prototypes for piano and how these prototypes are found to be in- effective in learning piano for the students due to the lack of expressiveness and improvisation in instrument learning. Most of these digitally augmented prototypes \cite{PIANO} provide continuous user feedback for the right and wrongly pressed keys while playing the instrument. One of the major disadvantages of these augmented reality prototypes is that they require an expensive learning curve to use and need an actual instrument (like a piano keyboard) on top which the the audio/visual feedback is provided through the AR technology. Additionally, most of the existing AR systems require bulky and expensive headsets that provide a limited field of view to the user playing the instrument. 
\subsection{Virtual Musical Instruments (VMIs) using gesture recognition and hand motion tracking}
With the popularization of image processing models and sophisticated motion tracking sensors, several existing technologies have emphasized on building virtual musical instruments using 3D hand gestures in-air \cite{motionsensor,motionsensor2,motionsensor4, motionsensor6, motion8} and/or Microsoft's Kinect 3D sensor\cite{kinect1,kinect2,kinect5,kinect6}. However, these sensor based virtual musical instruments are designed for playing instruments in-air, the experience of which is far from playing a real musical instrument. Additionally, some devices such as Kinect comprises of sensors such as 3D depth sensors, RGB cameras, and multiple microphones to interpret a 3D scene using continuously projected infrared light. While Kinect provides gesture recognition and body 3D motion recognition capabilities for the VMIs , these devices are highly expensive and their performance can deteriorate in the presence of outdoor lights and highly reflective surfaces. Additionally, the Kinect's motion tracking sensors do not perform well when the reaction is less than 0.5 seconds which is a plausible scenario while playing instruments \cite{kinectdis}. Similar to XBox technology, VMI using kinect sensor measures the RGB color and depth information followed by human coordinate measurement and gesture recognition. This triggers the corresponding sound from MIDI. Thus, these methods offer intangible solutions for instrument playing which may not be as fun and enjoyable to the user. 
Similarly, in \cite{rastogi_joshi}, the authors built a virtual musical instrument where a laptop camera is used to detect when a a stick hits their self-designed template, which then produces the associated sound via speakers. However, this system's robustness decreases when the user starts hitting the template with higher frequency. Additionally, there is a large latency overhead between the hit and sound generation which needs to be reduced for a real-time experience.  Moreover, one major caveat of using Kinect for playing instrument is that these systems  track only the position of skeleton joints which limits the playing experience of the user. It is important to track more detailed information w.r.t hitting intensity to play the corresponding sound from the instrument. Much like the AR/VR systems, some of the old Kinect sensors have a limited field of view, typically is less than 1 meter, which may not be suitable for playing very large and bulky instruments with a large surface area like piano. 

\subsection{Musical Instrument Apps and Games}
A wide-variety of musical instrument applications are available today for both Android and iOS users. These apps allow the users to learn and play a musical instrument on hand-held digital devices including smartphones or tablets. A few examples of such apps include Synthesia \cite{synthesia} for Piano, GuitarHero game \cite{guitarhero}, RealDrums \cite{RealDrums}, DrumPads \cite{drumpads}, and DrumKit \cite{drumkit}. While these apps are easily accessible from the app store, they are far from emulating a near real-time instrument playing experience. Additionally, a long-time usage of these mobile apps can lead to increase in screen time exposure which is can be detrimental in the cognitive development of the younger population \cite{cognitiveDevelopment}. Some user reviews for apps such as also reported complaints with respect to a considerable sound delay while tapping the drums too fast. Moreover, the interruptive in-app ads such as the reward ads tend to decrease the user engagement and retention while playing the musical instrument that lead to an overall negative user experience. 

Therefore, in this paper, we propose MuTable, which is a virtual, interactive instrument playing interface for playing Conga Drums.  To the best of our knowledge, none of the existing virtual musical platforms are designed for drums, which is a commonly played musical instrument. Typically, playing a drum-set requires strength and perfect eye-hand coordination which is not only complex but can also become a cumbersome task for beginners. To resolve this problem, we focus on designing this low-cost musical instrument platform for drums by projecting the instrument's surface area on a passive (without any sensors/actuators), tangible, and intuitive surface (such as table tops) to provide a unique instrument playing experience. The projected drum surface will simulate different binaural sounds that are essentially produced by the instrument using the hand gestures and taps performed by the user playing the drum kit. The proposed system can also serve as an excellent learning tool for the beginners who do not own an actual drum and can practise the correct hands movement, positioning, coordination, and timing using a portable, easy-to-access instrument interface. This  instrument interface can be easily setup by re-purposing any horizontal physical surface into a live musical instrument.

It is worth mentioning that most of the existing virtual musical platforms are inadequate in terms of either limited field of view (for AR/VR systems or Kinect sensors) or require pre-designed interactive surfaces with attached sensors underneath to facilitate a virtual instrument playing experience. Some existing works such as the VMI using Kinect sensors and MIDI have also reported a poor performance in terms of latency (\~ 11 seconds) which is not sufficient for a high quality instrument playing experience. Additionally, one common limitation of these existing works is that they do not consider modulating sound amplitude according to the user touch/tap intensity. This limits the user's playing behaviour and does not provide exposure to real-time music play for the user. To address these challenges, we design MuTable which allows re-purposing any flat/horizontal surface into a musical instrument of any kind. The MuTable system also incorporates different sound modulation by detecting different tap or touch intensities by the user. Moreover, it is a tangible solution where the user can have a complete field of view of the instrument to enjoy a close to real-time music playing experience.

Table 1 illustrates a summarized overview of existing Virtual musical instrument platforms and \textit{MuTable}.  
\begin{figure*}[!htb]
\centering
\includegraphics[width=0.9\linewidth] {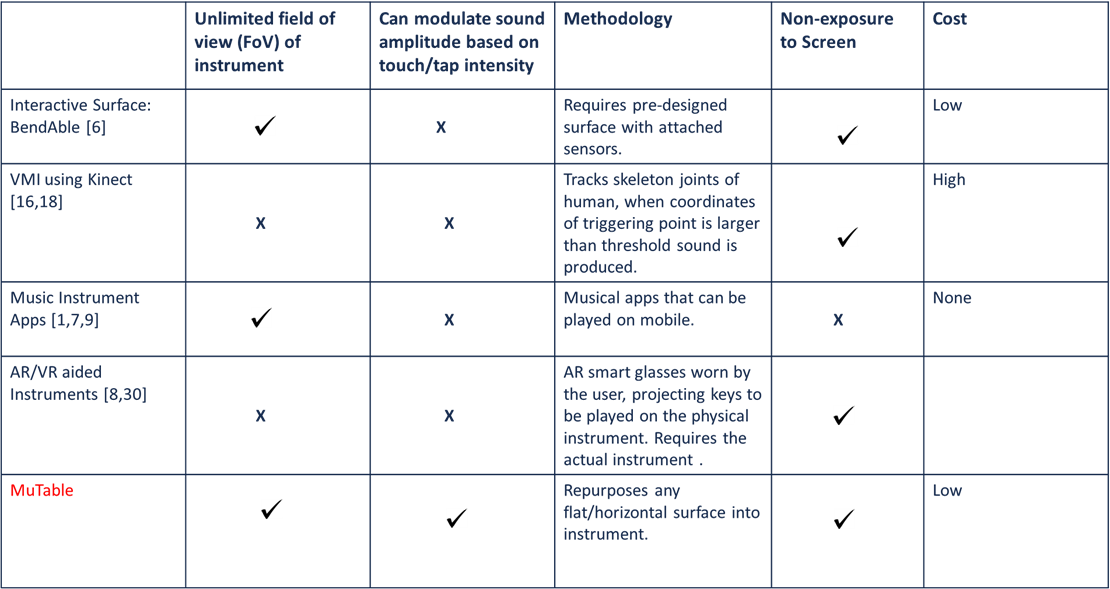}
\captionof{table}{Comparative analysis between MuTable and existing virtual musical platforms
}
\label{fig:Architecture}
\end{figure*}

\vspace{-1mm}
\section{Proposed Approach}
\label{sec:approach}
% approach(es) planned to solving problems - initial thoughts will suffice
\subsection{System Architecture}

Fig. 2 illustrates an end-to-end workflow of the proposed architecture. We discuss each component of this workflow in detail as follows:

\begin{figure*}
\centering
\includegraphics[width=0.9\linewidth] {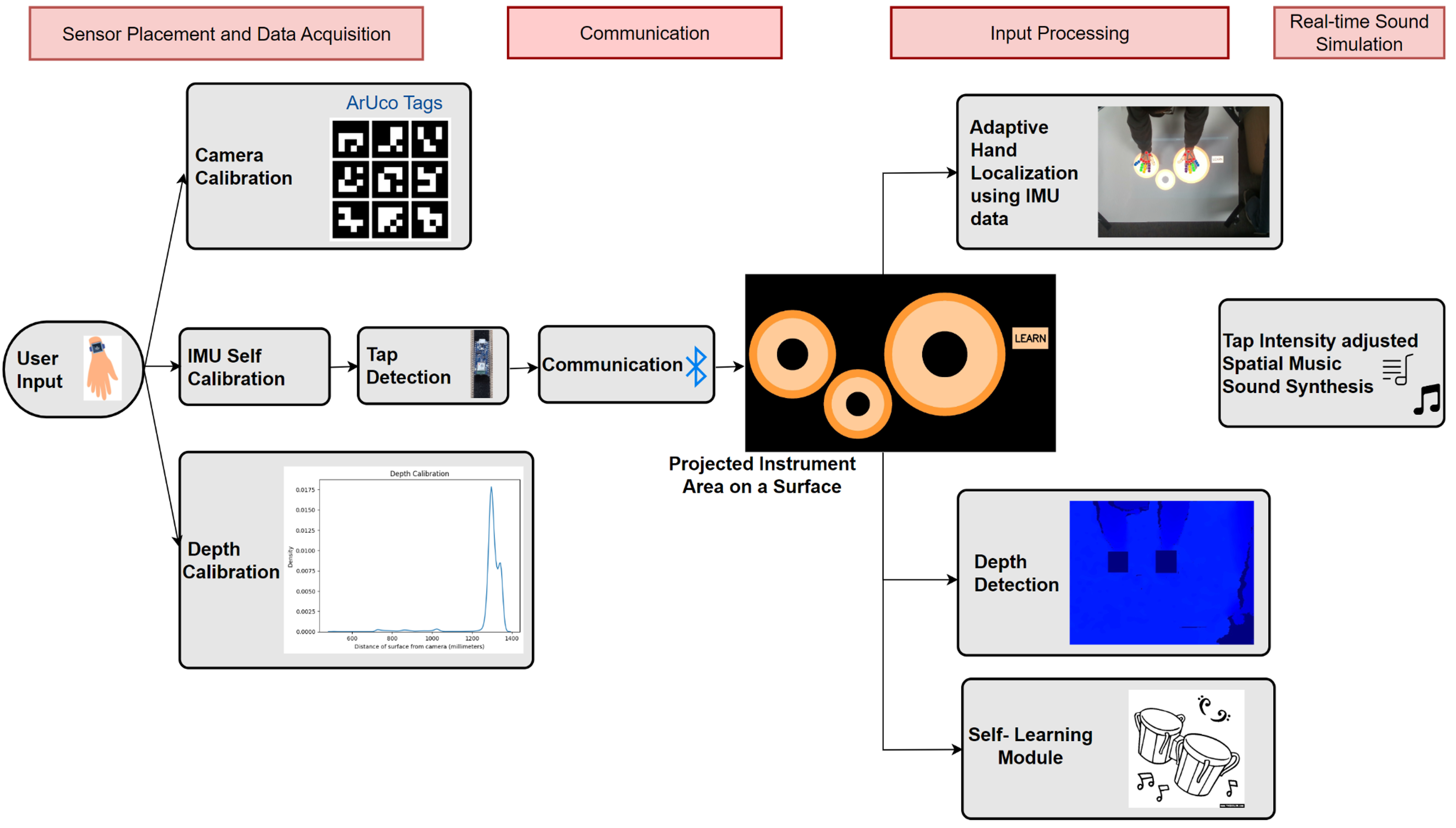}
\caption{End-to-end workflow
}
\label{fig:workflow}
\end{figure*}

\subsubsection{Hardware Architecture}
The hardware architecture comprises of a data acquisition module consisting of different sensors and a microcontroller to measure IMU data, capture hand gestures and detect taps. A centralized server is used for computational processing and a projector is required for projecting instrument surface.   
\begin{itemize}
    \item \textbf{Sensors:}  We used IMU sensors for tracking hand's tap detection using the accelerometer values.  We used the micro-controller's built-in IMU sensor to design wearable bands for tap detection. The output rate of the built-in accelerometer is set to 104 Hz. Additionally, a depth camera is used in for detecting the hand movement within the projected surface of the instrument. The depth cameras are directly connected to the centralized server to eliminate additional overhead w.r.t. communicating images over the network.

\item \textbf{High-end Microprocessor/Micro-controller:} Arduino Nano 33 BLE micro-controller is used as the edge device data acquisition unit for MuTable. The micro-controller has a powerful processor with 32-bit ARM® Cortex®-M4 CPU running at 64 MHz along with built-in Bluetooth module. The built-in 9 axis inertial measurement unit (including accelerometer, magnetometer and gyroscope) in Arduino Nano BLE, makes it a perfect choice for accurate tap detection as well as communication.

\item \textbf{Centralized Server:} A centralized server or a system is required to process the images from the depth cameras for adequate hand localization detection. The server used in our experiments is equipped with Intel Core i9-9820X CPU @3.30GHz.

\item \textbf{Additional Hardware:} A projector is required to project the desired instrument playing area on any flat/horizontal surface. 

\end{itemize}

 \subsubsection{Software Architecture}
The software architecture consists of a calibration module, tap detection module, input processing module and real-time sound simulation module. 
\begin{itemize}
    \item \textbf{Auto-Calibration Module:} Sensor placement and self calibration is highly important step for accurate detection of taps and hand localization in the MuTable system. To ensure the sensors' accuracy and repeatability, we perform the calibration process apriori before the user starts playing the virtual instrument. Each of the sensing unit is calibrated independently. For instance, the camera module is calibrated using the ArUco Tags \cite{aruco}. We measure the confidence of the calibration based on the number of detected markers. If it is above the predefined threshold, then the musical instrument is projected and the system is ready to use. The depth sensor is auto calibrated using the depth estimates of point cloud at start of the session. Fig. 3 illustrates the probability density function of depth of the detected points. As the camera view is parallel to the projected surface, the most frequent depth estimate is assumed to be the surface depth.
    
    \begin{figure}[!htb]
\centering
\includegraphics[width=\linewidth] {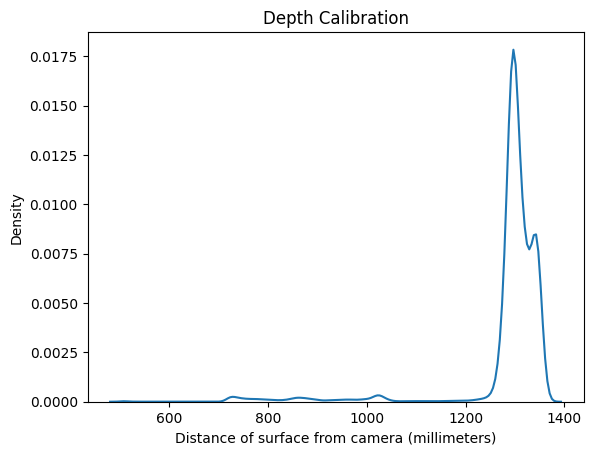}
\caption{Probability density of depth of point cloud}
\end{figure}

    For IMU sensor calibration, the user with the wearable band is required to tap the surface a few times before playing the instrument on the projected area. This allows the IMU calibration module to automatically set the acceleration thresholds for tap detection. The differential change in acceleration values is different for different sensors and Arduinos. Therefore, to detect the differential change co-responding to a tap for a given sensor/arduino this self-calibration module is programmed within the Arduino during the setup phase.

\item \textbf{Tap Detection Module:} A tap, in this context, is an action when the user taps his/her palm on the projected drum surface. Since an accelerometer is sensitive enough to detect slight changes in the hand motion due to a nearby tap, therefore a tap event can be detected by the accelerometer as an instantaneous acceleration shock or jerk. To detect tap using acceleration we monitor how fast the hand is moving in the up-down direction.  Therefore, the tap detection algorithm analyzes the acceleration data from the tri-axes accelerometer in near real-time. In order to determine the instantaneous tap event, we compute the rate of change of acceleration (or tap intensity). In other words, a time derivative for the acceleration signal along a given axes (Z-axis) is calculated periodically . A tap is determined if the calculated acceleration derivative exceeds a predetermined threshold value for a specific interval of time. 
 Mathematically, taps for a given time series of IMU data can be calculated as follows, 
\begin{figure}[!htb]
\centering
\includegraphics[width=0.55\linewidth] {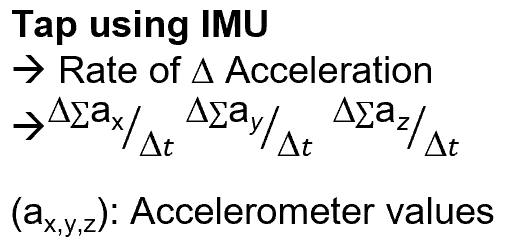}
\label{fig:equation}
\end{figure}

The threshold value is calculated using the training samples collected in during the calibration phase of the system.  

However, thresholding the sensor readings alone was inadequate in this system because it produced a high rate of false-positive alarms, wherein consecutive taps were being reported for a single-tap event. This was because of the high output rate of the Nano BLE IMU sensor (104 Hz) such that the instantaneous impact of a tap was much longer than the output rate of the sensor resulting in false alarms.  In order to address this problem, we augmented the tap-detection algorithm with time-based threshold technique to filter out consecutive tap events within a predetermined small period of time. 

\item \textbf{Multi-modal Input Processing Module:} The input processing module consists of two sub-components; 1) Adaptive hand localization using two different modalities of IMU data and camera images 2) Depth detection. Precise hand location detection on the projected instrument surface is important to produce sounds depending on the instrument shape that was tapped on. Deep learning based image
segmentation technique is used for hand localization
from real-time images streamed from the depth camera \cite{mediapipe}. Along with detecting taps in the z-direction, IMU sensor tracks the motion of the hand in the music instrument projected plane. IMU sensor sends the tap event to the centralized server along with the flag indicating if there is any movement in the projected plane. On receiving the tap event, if the flag indicates the significant movement in the projected plane, camera images are processed using the palm detection model followed by the hand landmark model (single-shot model as discussed in \cite{mediapipe}) to identify the precise location of hand placement on the projected area of the instrument. If the hand is placed outside the bounds of the instrument surface, the tap event will be ignored by the server. 

Once the hand location is detected on the projected area, the depth sensing module measures if the hand is close to the projected plane. This is important to check to avoid the false positives when the user's hands are not close to the surface. If it passes this check, a sound is produced associated to the location and instrument shape on which the tap event was performed by user. Fig. 4 illustrates hand images with ground truth annotation (left) and depth detection hand placement with respect to the projected surface (right).

\begin{figure}[!htb]
\centering
\includegraphics[width=\linewidth] {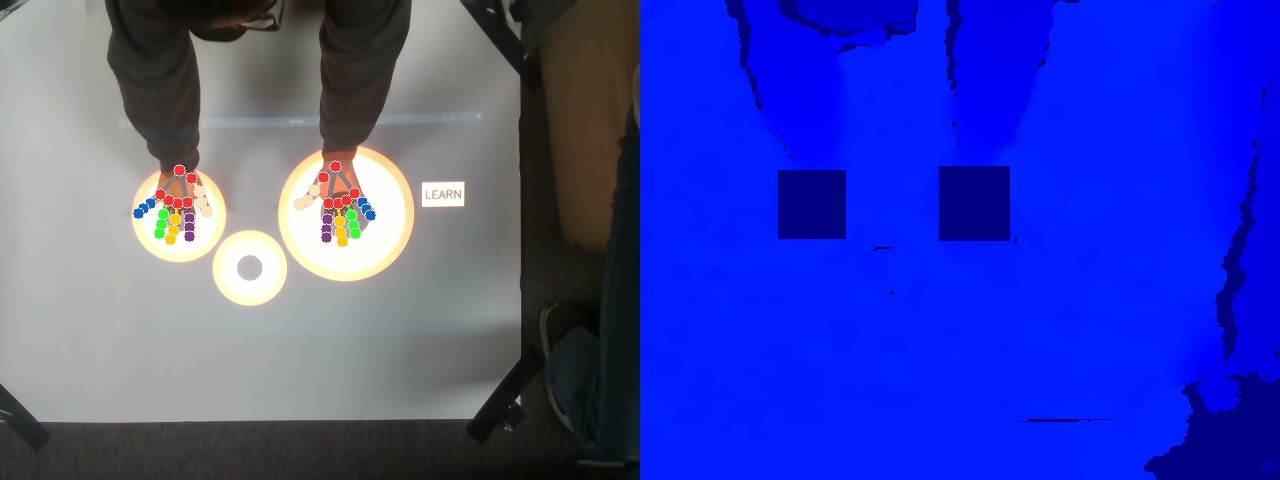}
\caption{Hand localization (left) and depth detection (right)
}

\label{fig:setup}
\end{figure}

\item \textbf{Self-Learning Module:} The Self-Learning module displays a pre-designed composition on the projected instrument surface to help users in their instrument learning process.  
\item \textbf{Real-time Sound Simulation Module:} A sound bank is synthesized on the server to generate real-time spatial sounds with different amplitudes based on the tap intensity detected and sent by the IMU sensor. 
\end{itemize}

 \subsubsection{Communication}
The built-in Bluetooth module of Arduino Nano BLE is used for sending acceleration data to the server whenever a tap is detected.

\section{Implementation}
Fig. 5 illustrates two core components of the experimental setup a) a wearable band with Arduino Nano and b) Projector attached with a depth camera. 
Fig. 6. illustrates the overall system setup with depth camera and projector mounted on a stand, and a projection of three Conga drum discs surface on the ground. The GitHub repo for the \textit{MuTable} code is available publicly \footnote{https://github.com/akash17mittal/muTable}

\label{sec:ExperimentalSetup}
\begin{figure}[!htb]
\centering
\includegraphics[width=\linewidth] {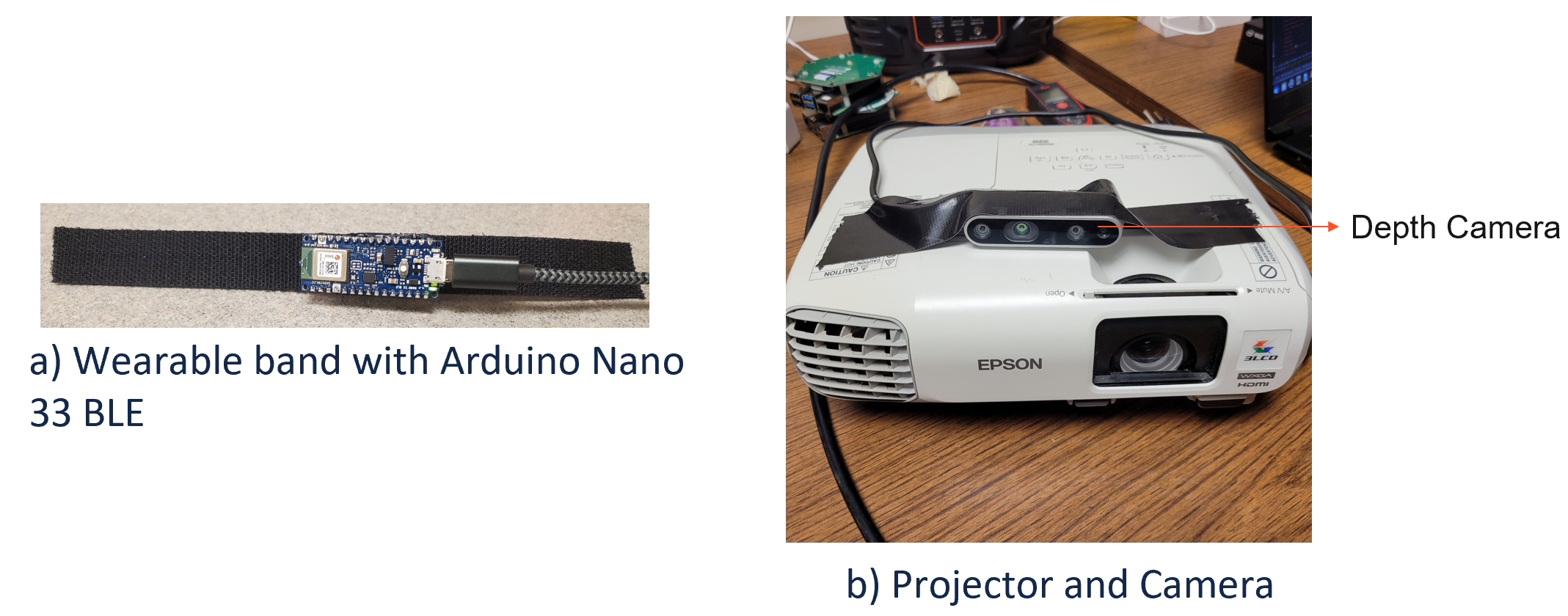}
\caption{Experimental Setup-1
}

\label{fig:setup}
\end{figure}

\label{sec:ExperimentalSetup}
\begin{figure}[!htb]
\centering
\includegraphics[width=0.55\linewidth] {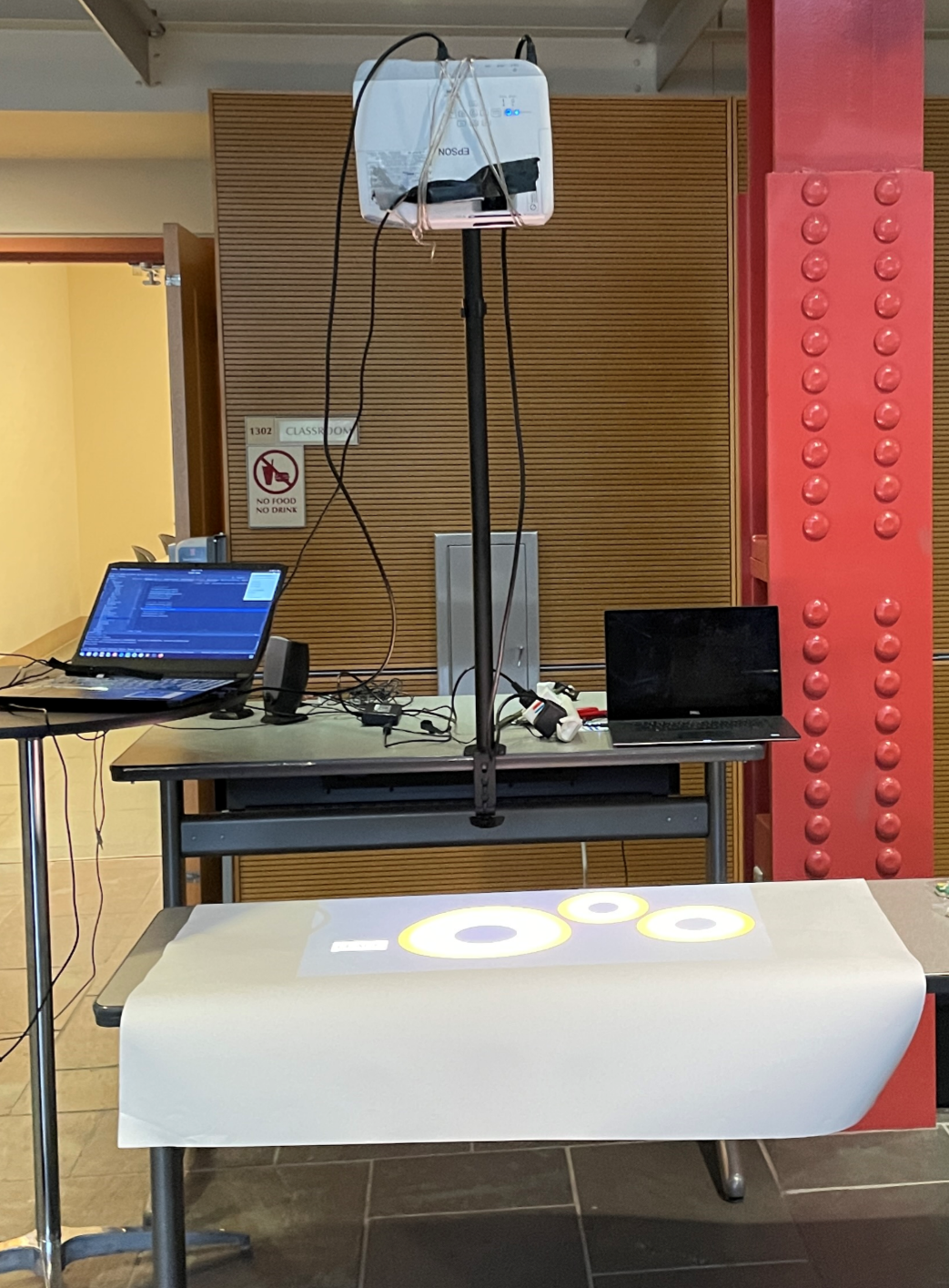}
\caption{Experimental Setup-2
}

\label{fig:setup}
\end{figure}

\section{Evaluation}
To evaluate the proposed system, we will assess the system using two quantitative metrics (accuracy and latency) and one qualitative study. 
\subsection{Quantitative Evaluation}
The quantitative metrics include the end-to-end latency, i.e. the time elapsed between the tap and the generated sound, and accuracy of the tap detection and hand localization. We demonstrate results for each of sub-component of the MuTable system:
\subsubsection{Tap Detection}
Accurate tap detection is an integral part of the MuTable. Fig. 7. illustrates a graph for acceleration derivative (tap intensity) signal with respect to the number of samples. As shown in the Fig. 7 a), there is a sudden fall in the  when a tap is detected. Fig. 7 b) shows a close-up view of the IMU signals for a subset of samples when a tap is detection. 
As evident from the figure, whenever the tap intensity value (along Z axis) drops below -0.5,  a tap is detected. To prevent false alarms with consecutive reporting of taps for a single tap event, we set the time-based threshold to 500 ms (higher than the IMU sensor's output rate of 104Hz or 
9 ms).  

While our tap detection algorithm gave accurate results with almost 100\% accuracy for hard taps (which is usually the case with playing drums), we also evaluated the algorithm's performance in the presence of softer taps (or taps with smaller intensity). This is because the MuTable system is programmed to generate different amplitudes of drum sounds depending on the varying tap intensities and instrument shape. Fig. 8. illustrates a confusion matrix for the tap detection algorithm in the event of softer taps. As evident from the confusion matrix, 9 False Negative alarms are detected in this experiment reducing the accuracy of tap detection to 83\%. With further exploration, we observed that these false negative alarms were generated only when the change in tap intensity is very small. This is acceptable in our case because it helps in avoiding detection of random soft hits on the surface as taps.   

\begin{figure*}[!htb]
\centering
\includegraphics[width=\linewidth] {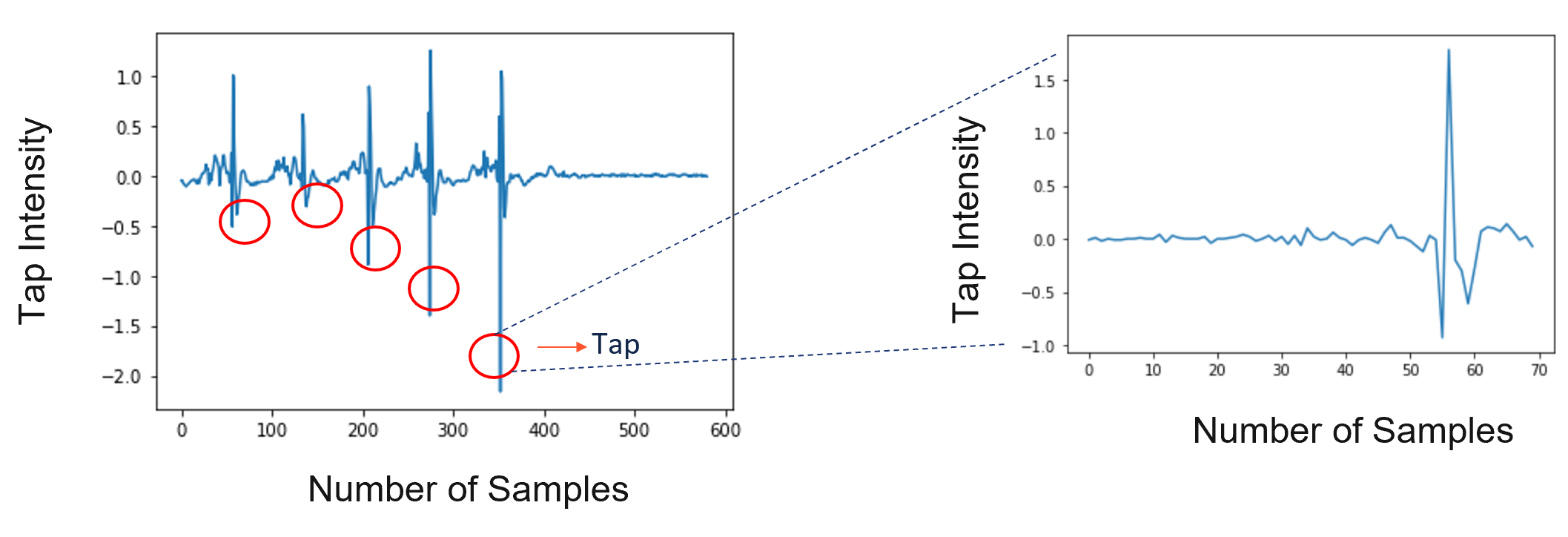}
\caption{a) Acceleration Derivative along X-axis (Tap Intensity) Vs. Number of Samples b) A close-up view of the signal when a tap is detected instantaneously
}
\label{fig:Workflow}
\end{figure*}

\begin{figure}[!htb]
\centering
\includegraphics[width=\linewidth] {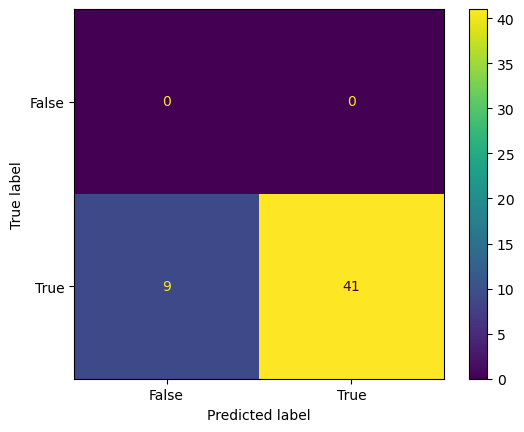}
\caption{Confusion Matrix for Detecting Softer Taps (taps with small intensity)
}
\label{fig:Workflow}
\end{figure}

\subsubsection{Latency}
We computed latency across three dimensions; Bluetooth communication latency from Arduino Nano to server, latency for detecting hand localization, and end-to-end latency. The end-to-end latency is the aggregated sum of communication latency and hand-localization latency. The average BLE communication latency between Arduino Nano and Server is measured as 100.3 ms for sending tap events as binary data of size 24 Bytes. Similarly, the average latency for hand localization completion time on the server side is measured as 24 ms. Therefore, the average end-to-end latency of the system is observed as 124.3 ms.  

 \textbf{Optimized Camera Processing:}

 In order to optimize the redundant computations on the image data from camera, we used the IMU data to trigger camera based processing only when it is required. In other words, we tried to optimize computational latency for images by only processing the images from the camera when a hand is moved by the user in the X-Y plane of the projected surface. This is because intuitively,  if the user continues to tap on the same location of the projected surface repeatedly, image processing for the event with same spatial locality will be redundant. Therefore, to minimize this overhead, the edge device (Arduino Nano)  detects change in acceleration along the 2D (X-Y) plane of the surface, and whenever the hand moves along the 2D plane it will trigger the depth camera to activate hand localization process. Fig. 9. illustrates a graph of IMU signal with change in acceleration values for hand movement along the 2D plane of the projected surface. The signal depicts that as the user moves further away from the initial tap  location, the change in 
 acceleration value also increases. This change in acceleration values along the Y and X axes corroborate that the user has tapped on a different spatial location or on a different drum shape in comparison to the previous tap instance. This activates the image based processing from the live camera stream to produce the associated sound. 

 Fig. 10 shows a graph of latency for BLE, hand localization and end-to-end system in two different setups, i.e. original setup with continuous image processing and optimized image processing in the event of consecutive taps with same spatial locality. As observed in the figure, the end-to-end latency in the optimized setup goes down to 100.3 ms which is the same as the latency for Bluetooth communication. This is because in the optimized setup by using the spatial locality from IMU data, the overhead for hand localization latency can be eliminated (i.e. hand localization latency becomes zero when user's consecutive taps are on the same planar locality).

 \textbf{Bluetooth Performance w.r.t data:}
We experimented with two different modes of data transfer between Arduino Nano and server; first, for sending binary data for tap events (of size 24 Bytes) and second, for sending raw acceleration values as String data (of size 62 Bytes). 
 Fig. 11 demonstrates the latency distribution for sending data from Arduino Nano BLE to server with a varying size of 24 Bytes and 62 Bytes. As evident from the plot, for sending 24 bytes of data via BLE, the average latency was around 100.3 ms whereas the average latency for sending 62 Bytes of data was around 110.87 ms. It is worth mentioning that as we increase the data size from 24 to 64 Bytes, the average latency increases with an additional overhead of 10.57 ms. It is worth mentioning that while our basic setup was used to send binary data (0 or 1 for tap events), in the later stage, we had to send raw acceleration values to produce sound signals on the server based on the tap intensity measured by the IMU sensor.   

 \textbf{Discretization:} Range based data discretization technique is used to reduce the amount of data being sent from the edge device to the server. Since the binary data (0 or 1 for tap) cannot preserve knowledge about the tap intensity, we applied range-based discretization technique on the continuous IMU sensor values. The number of intervals and the range for each interval during the discretization process is determined based on the different amplitudes of sound that the server should generate corresponding to the tap intensity. This method is advantageous and tries to overcome the limitation of sending continuous float acceleration values while preserving the notion of tap intensity which is not possible in binary data. Fig. 12 shows the 
 discretized graph for acceleration signal along Z-axis, that is used to detect a tap with varying intensity within each interval/bin. 
\subsubsection{Hand Localization Accuracy and Latency:}
By applying the single shot model \cite{mediapipe} constituting of both the palm detection model and hand landmark model, we achieve an average precision of 95.7\%.   . 

Fig. 13, illustrates the latency distribution of hand location detection. The average latency for hand localization detection is measured to be 24 ms. 

\begin{figure}[!htb]
\centering
\includegraphics[width=\linewidth] {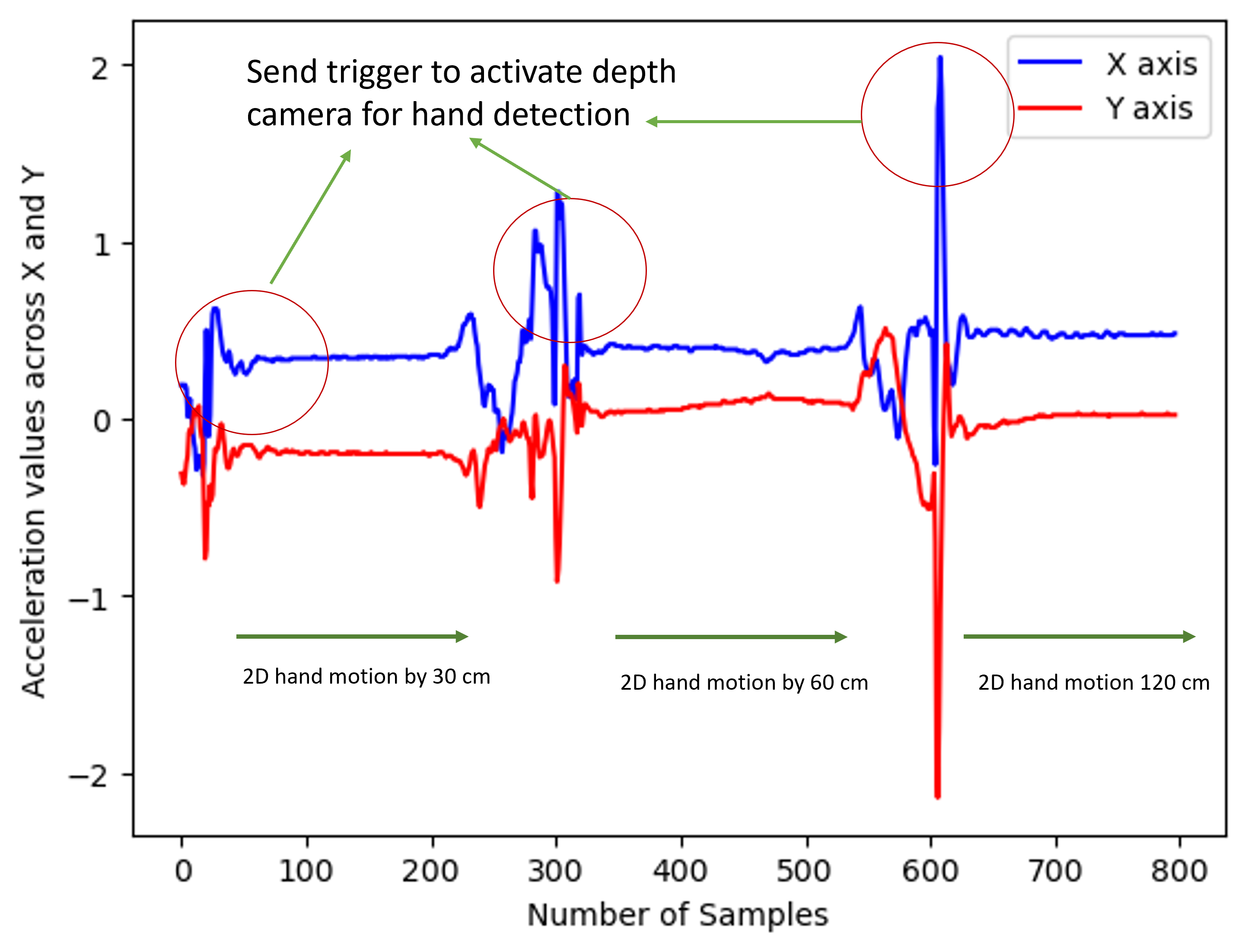}
\caption{Change in Acceleration Values for Hand Motion across 2D (X-Y) plane
}
\label{fig:optimizedCommunication}
\end{figure}

\begin{figure}[!htb]
\centering
\includegraphics[width=\linewidth] {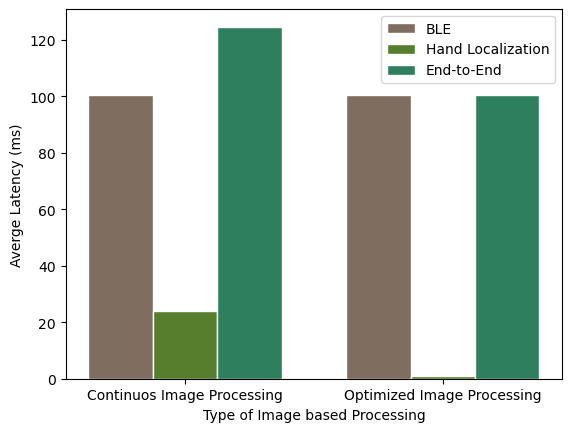}
\caption{Average latency for continued image processing and optimized image processing for taps with same spatial locality
}
\label{fig:}
\end{figure}

\begin{figure}[!htb]
\centering
\begin{subfigure}[Arduino BLE Transfer rate for 24 Bytes]
{\includegraphics[width=0.46\linewidth] {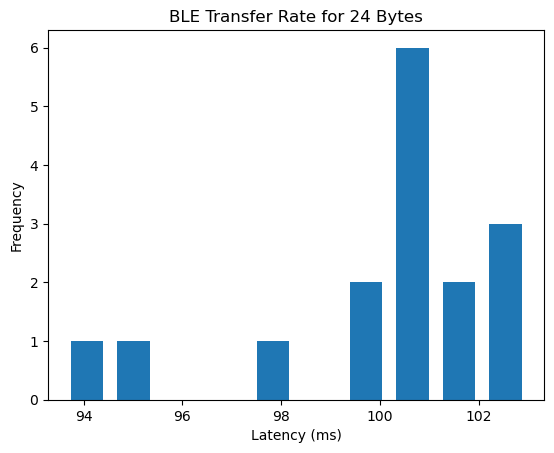}
  \label{fig:Ble24Bytes}
 }%
\end{subfigure}
\begin{subfigure}[Arduino BLE Transfer rate for 62 Bytes]
{\includegraphics[width=0.46\linewidth] {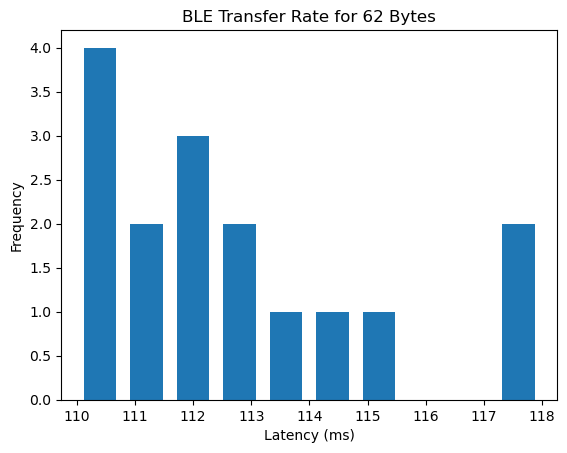}
  \label{fig:Ble64bytes}
 }%
\end{subfigure}
\caption{Transfer Rate for Arduino BLE to Server}
\label{fig:BLEDistribution}
\end{figure}

\begin{figure}[!htb]
\centering
\includegraphics[width=\linewidth] {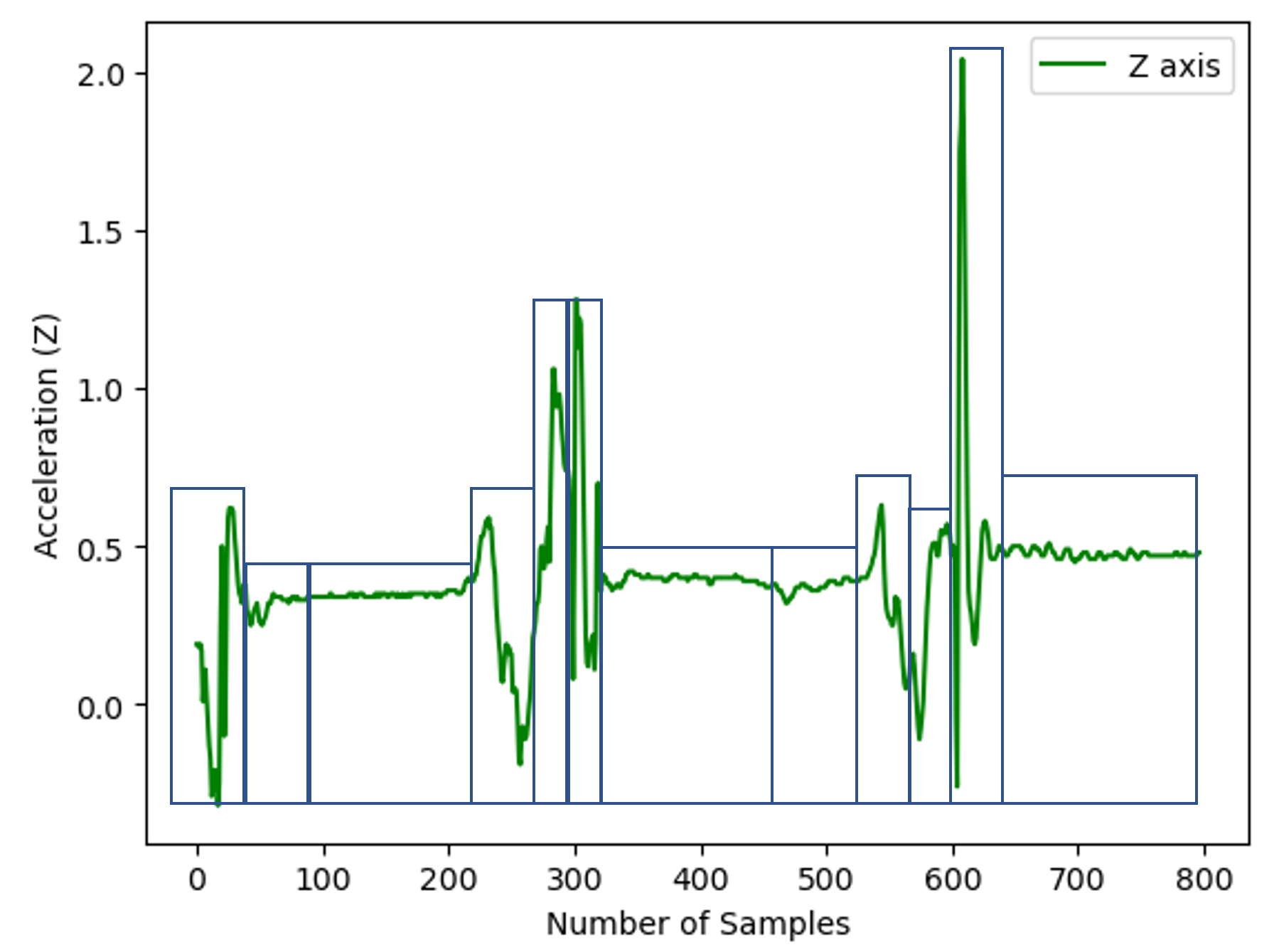}
\caption{IMU Sensor Data Discretization
}
\label{fig:handlocalizationLatenc}
\end{figure}

\begin{figure}[!htb]
\centering
\includegraphics[width=\linewidth] {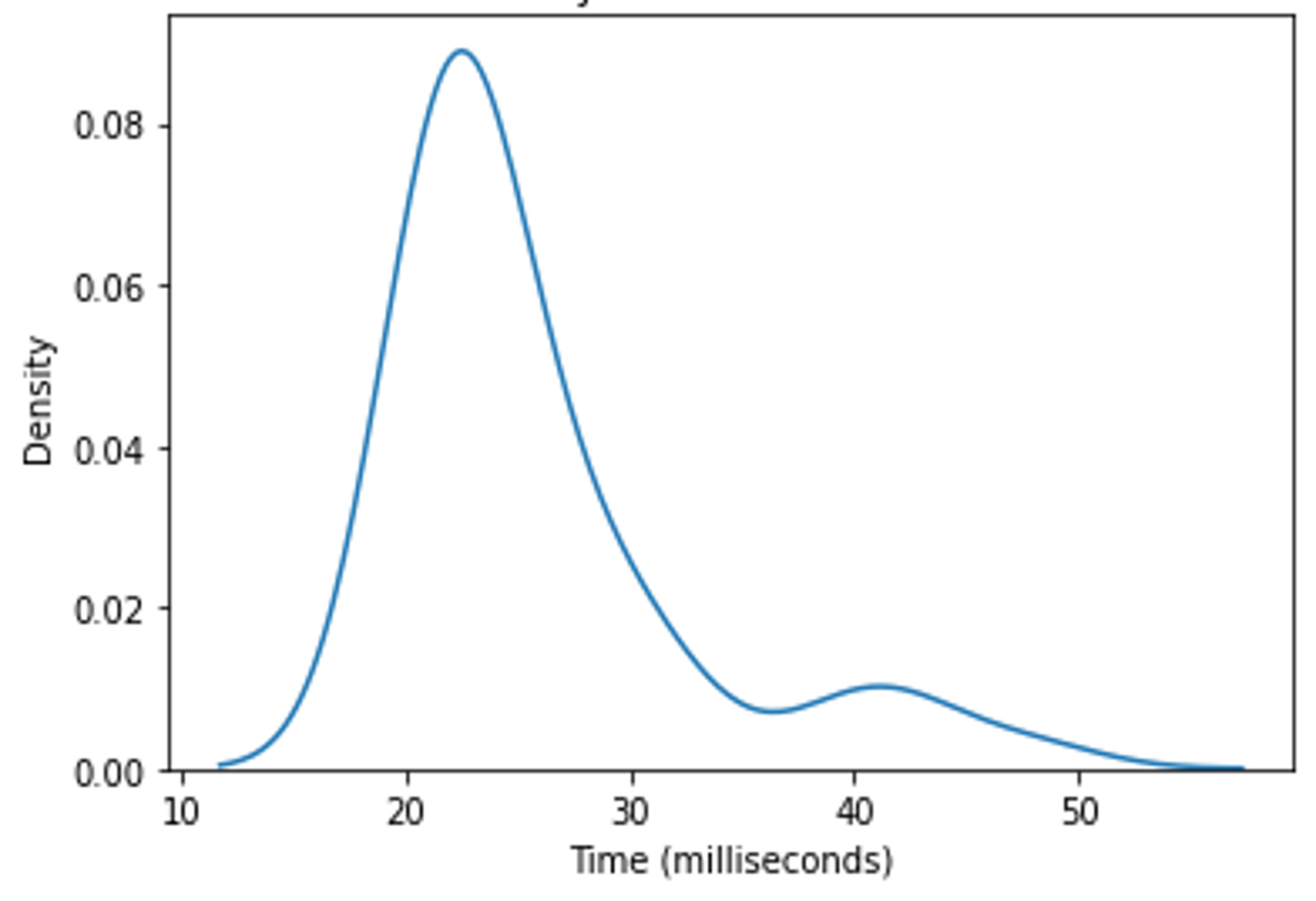}
\caption{Probability Latency for Latency Distribution of Hand Location Detection
}
\label{fig:handlocalizationLatenc}
\end{figure}

\subsection{Qualitative Study}
 For qualitative assessment, we estimated Quality of Experience (QoE) by conducting user studies with three different users' playing MuTable.  QoE is an important metric for our work in order to analyze how close this virtual interactive experience will be to a real instrument playing experience.

\begin{itemize}
    \item \textbf{User Study-1:}
    \url{https://youtu.be/jDGnFMR6C9o}
\item \textbf{User Study-2:}
\url{https://youtu.be/N4G_yXEIZEw}
\item \textbf{User Study-3:}
\url{https://youtu.be/jGTT_nC8-Ss}
 \item \textbf{User Study-4:}
\url{https://youtu.be/vaPYGNbSFbo}

 \item \textbf{User Study-5:} This was done to study the self-learning module of MuTable. 
\url{https://youtu.be/4biQIoIg3gA}
\end{itemize}
\section{Discussion}

\textit{MuTable} is a real time system that requires minimal latency when the user taps and the associated sound is synthesized. To replicate a similar experience as of playing a real music instrument, we made specific design choices and did several optimizations as follows:

\begin{itemize}
    \item \textbf{On-device Tap Detection:} The tap detection algorithm is implemented on-device i.e. on the motion sensing band itself. This has several benefits as it saves the communication cost of sending raw sensor data to the server and it is fast too as the computation is done closer to the data. \\
    \item \textbf{Adaptive Hand Localization:} Processing the image for localising hands takes about 25 ms. To reduce this time, camera images are processed adaptively depending on the hand movement in the music instrument projected plane. If the motion sensing bands don't detect movement in the projected plane, the image processing step is skipped and the hand location is assumed to be the same as of previous tap. \\ 
    
    Additionally, we observed blurry images when user taps frequently. This affects the performance of hand location detection. This adaptive hand localization technique effectively mitigates this problem. \\

    \item \textbf{Auto-calibration:} All the sensors are auto-calibrated that frees the user from specifying the parameters manually such as the height at which the projector and camera are mounted, tap detection threshold, etc. User can freely change the location of the system without worrying about calibrating sensors. As different users have different tapping patterns in terms of intensity, the system automatically adapts to multiple users. \\
    \item \textbf{Sound intensity proportional to tap intensity:} To provide the similar experience as of playing a real instrument, motion sensing band communicates the intensity of tap to the server. The amplitude of the synthesized sound is adjusted according to the tap intensity. Additionally, the spatial sounds are synthesized depending on the placement of different components of the music instrument by adjusting the amplitudes of left and right channel of the synthesized sound. \\
    
    \item \textbf{Reducing false positives: } The number of false positives is very common for such systems as any random hand gesture or slight tap on the table can be detected by the system to produce the corresponding drum sound based on the hand location. Therefore, it is important to reduce the number of such false positives. To solve this problem, we chose to do apriori calibration of the sensors and determine the intensities of tap specific to the user at the start of the session. \\
    
    Additionally, we observed a number of false-positive alarms for our tap detection algorithm wherein multiple taps were being reported for a single-tap event. In order to address this problem, we augmented the tap-detection algorithm with time-based threshold technique to filter out consecutive tap events within a predetermined small period of time.  \\
\end{itemize}

\section{Future Work}

In the current work, we utilized multiple sensors including IMU sensor, camera, and depth sensor. This work can be further extended by relying on the fewer sensors. Possible directions can be:
\begin{itemize}
    \item \textbf{Removing the depth sensor:} The depth can be estimated based on the size of the detected hand. At the start of the session, user can be asked to place hands on the music instrument projected surface. The size of the detected hand acts as a base for the depth estimation. As the user moves the hand in the vertical direction, the size of the detected hand will vary in the camera view. This can be utilized to estimate of depth at which the user taps.
    \item \textbf{Fine grained finger movement detection:} To extend \textit{MuTable} to more music instruments like piano, more precise finger movement tracking is required. The current motion sensing band is worn on wrist that limits its capability to track fine grained hand movements. Optimal placement of motion sensor can be found combined with the camera sensor to detect the hand and finger movements beyond taps.
    \item \textbf {Overcoming communication latency: } There will always be some communication cost of sending taps to the server that is a bottleneck. It can affect user experience too. To address this, we can introduce intuition based learning i.e., predicting the tap based on the current tapping pattern for e.g. continuous tapping. Sounds can be produced even before the band communicates the tap to the system.
\end{itemize}

\bibliographystyle{acm}
\bibliography{ref}

\end{document}